%% file: paper.tex
\documentclass[twocolumn]{article}
\usepackage{commentstar}
\usepackage{epsfig}
\usepackage{amssymb}
\usepackage{amsmath}

\makeatletter
\DeclareRobustCommand{\qed}{%
  \ifmmode 
  \else \leavevmode\unskip\penalty9999 \hbox{}\nobreak\hfill
  \fi
  \quad\hbox{\qedsymbol}}
\newcommand{\openbox}{\leavevmode
  \hbox to.77778em{%
  \hfil\vrule
  \vbox to.675em{\hrule width.6em\vfil\hrule}%
  \vrule\hfil}}
\newcommand{\qedsymbol}{\openbox}
\newenvironment{proof}[1][\proofname]{\par
  \normalfont
  \topsep6\p@\@plus6\p@ \trivlist
  \item[\hskip\labelsep\itshape
    #1.]\ignorespaces
}{%
  \qed\endtrivlist
}
\newcommand{\proofname}{Proof}
\makeatother
\usepackage[dvips]{color}

\makeatother

\newtheorem{theorem}{Theorem}
\newtheorem{corollary}{Corollary}

\newtheorem{lemma}{Lemma}

\begin{document}
\title{Lower bounds for distributed \\ Markov chain problems}
\author{Rahul Sami \\ School of Information \\ University of Michigan
  \\ \texttt{rsami@umich.edu} \and Andrew Twigg \\ Computing Laboratory
  \\ University of Oxford \\ \texttt{andy.twigg@comlab.ox.ac.uk}
}
\maketitle

\begin{abstract}
We study the worst-case communication complexity of distributed algorithms computing a path problem based on stationary distributions of random walks in a network $G$ with the caveat that $G$ is also the communication network. The problem is a natural generalization of shortest path lengths to expected path lengths, and represents a model used in many practical applications such as pagerank and eigentrust as well as other problems involving Markov chains defined by networks.

For the problem of computing a single stationary probability, we prove an $\Omega(n^2 \log n)$ bits lower bound; the trivial centralized algorithm
costs $O(n^3)$ bits and no known algorithm beats this. We also prove lower bounds for the related problems of approximately computing the stationary probabilities, computing only the ranking of the nodes, and computing the node with maximal rank. As a corollary, we obtain lower bounds for labelling schemes for the hitting time between two nodes.
\end{abstract}

\input{markov_chains_lowerbounds}

\end{document}

%% file: markov_chains_lowerbounds
\section{Introduction}
\label{section:introduction}

Let $G$ be a strongly-connected, directed, unweighted graph on $n$ vertices. $G$ defines a communication network, where nodes are processors and communication can only occur along edges. $G$ also defines a {\em random walk} process: a token walks over the nodes (states) of $G$, at each time step choosing the next node uniformly at random from the outgoing edges of the current node. We shall refer to this stochastic process as the {\em harmonic random walk on $G$}.

A basic path problem in distributed computing is as follows. Given a network $G$ where each node only knows its neighbours, compute the lengths $d(v,u)$ of the shortest path in $G$ from each node $v$ to some fixed node $u$. We consider the natural generalization of this problem to expected path lengths using the harmonic random walk on $G$. Let $E[d(u,v)]$ be the expected length of the walk that begins at node $u$ and terminates on first hitting node $v$. We shall be interested in the values $E[d(u,u)]$, i.e. the expected return times of the token. If the Markov chain defined by $G$ is irreducible then these values exist and the random walk has a unique {\em stationary distribution} $\pi=(\pi(1),\pi(2),\ldots,\pi(n))$ where $\pi(u) = 1 / E[d(u,u)]$ is known as the {\em stationary probability} of node $u$ and $\sum_u \pi(u) = 1$.

If $G$ is undirected then the stationary probability of any particular vertex is proportional to its degree -- regardless of the structure of $G$. This remarkable fact is the key to numerous applications involving Markov chains. Crucially however, the networks we consider are directed and so in general, no such simple closed-form expression exists for $\pi$.

Random walks have been studied extensively and have numerous applications in distributed computing including self-stabilizing networks and token management \cite{coppersmith93}. In the last decade there has been substantial interest in using random walks to construct ranking schemes in networks, the most obvious being the pagerank scheme used by Google to rank web pages \cite{page98pagerank}.

In this paper we consider distributed algorithms that compute properties of the harmonic random walk on some network $G$, with the caveat that the algorithms must use $G$ both as a communication network and as an input, where each node initially knows only its local edges.

Fix some node $u$. We say a distributed algorithm {\em computes} a value $x$ iff it
terminates with at least one node knowing $x$. The {\em communication
complexity} of a distributed algorithm is the total number of bits sent,
over all edges. The communication complexity of computing a value $x$ is
the minimum communication complexity of any distributed algorithm that
computes $x$. Our main aim is to show good lower bounds on the communication complexity of distributed algorithms computing $\pi(u)$, and some related problems.
Note that, since the harmonic random walk uses only rational probabilities (the reciprocal of the outdegree of a node), the stationary probabilities are also rational and can be computed using finite precision. Our problem, then, is not how to efficiently approximate a real number, but how to understand its inherent complexity based on the network topology that generates it.

Our main result is a series of lower bounds that suggest that, in the worst-case, one can do better than the trivial centralized algorithm, and in some cases randomization may be of help in reducing the communication complexity.

\subsection{Technique}
\label{sec:technique}

Our main technique for proving lower bounds is to consider a related two-party
communication complexity problem: partition the nodes of $G$ into $G_1,G_2$ and let Alice know $G_1$ and Bob know $G_2$, and choose some node $u \in G$. We first lower bound the number of bits Alice and Bob must exchange in order to compute $\pi(u)$ in the worst-case (see \cite{nisan97communication} for examples of this technique). Then we lift the result to the distributed case by replacing the cut $\left< G_1,G_2 \right>$ by a linear array, `stretching' each edge of the cut by as many edges as possible while maintaining $O(n)$ nodes in the network. Because of this, many of our worst-case instances resemble the `barbell graph'. To do this we appeal to the linear array conjecture \cite{237817,258622} as follows.

If we can show that there exists a class of graphs having a cut where at least $\Omega(n)$ bits must be communicated across this cut, and if the cut is sparse (contains $O(1)$ edges), we can replace it by a line of $n$ edges. The question `does this increase the communication complexity by a factor $n$?' is the linear array conjecture \cite{237817}. The answer is that the randomized communication complexity increases by a factor $kn$, where $k$ is some constant less than $1$. In other words, each of these edges must see $\Omega(n)$ bits.

We shall prove all our two-party lower bounds by reduction from two main known problems. For the purely information-theoretic bounds, we reduce from {\em set-disjointness}: Alice and Bob each have a subset $P,Q$ of $\{1,\ldots,m\}$ and they must decide whether $P \cap Q = \emptyset$. The randomized communication complexity of disjointness is $\Omega(m)$ bits, for any protocol that decides with error probability less than $1/3$. Some of our results give bounds much stronger than the information theoretic results, but only for deterministic algorithms. For these results we use the problem {\em greater-than}: Alice and Bob each have an $m$-bit number $P,Q$ and they must decide whether $P \geq Q$. Any deterministic protocol must have communication complexity $\Omega(m)$ bits, yet any randomized protocol must communicate at least $\Omega(\log n)$ bits. Other results and proofs in communication complexity can be found in the book \cite{nisan97communication}.

Our technique resembles that of Tiwari \cite{32978}, where the network is simulated on a linear array and then one can use a reduction from a known bound on the complexity of the function when computed on a linear array of processors. Our work differs from Tiwari's however, since we require that algorithm must use the network both for communication between processors and as the input. For this reason we cannot consider the function to be computed and the network that it is to be computed on as two separate problems. In particular, there appears to be a tradeoff between the strength of a two-party lower bound and our ability to `lift' it onto a larger network to obtain good bounds in the distributed case. In this sense, our problem is similar in spirit to that of Hakowiak et al. \cite{314705} who consider the problem of distributedly computing maximal matchings of a network (although they consider upper bounds on the time complexity rather than communication).

We believe that strengthening our lower bounds requires a different technique. This is because we believe that the two-party lower bounds are tight if the cut is sparse, but the lifting onto a linear array only amplifies the result by a factor of $O(n)$. In other words, the two party bound is replicated across at most a {\em linear} number of edges, and yet there are potentially $O(n^2)$ edges in such a network of low expansion and high diameter.

\subsection{Related work}

Aside from being an interesting problem in its own right, the problem is an
abstract model underlying several basic problems in distributed computing, for
example randomized routing, self-stabilization, network flow \cite{48689} and load balancing (where a node offloads work to its neighbours in proportion to their difference in current workloads). Although distributed and parallel algorithms have been used to solve Markov chains simulating large systems, for example queuing or communication networks, our setting is different -- the Markov chain that we wish to sample from is defined precisely by the communication network on which it needs to be computed.

The problem also underlies the pagerank algorithm \cite{page98pagerank} for web page ranking. Here, $G$ represents the web graph where each node is a web page and an edge represents a hyperlink between two pages. The pagerank of $G$ is defined as the stationary distribution of the harmonic random walk on $G'$, where $G'$ is obtained by adding a `reset' transition from every node to every node in some root set $S$. In a problem of this scale, distributed computing in the network could be extremely valuable
\footnote{For pagerank, the nodes do not map directly to network 
nodes, as there are usually many pages on a single site. However, even 
if we collapse all pages on a site to a single page and abstract to the level
of network nodes, the problem is still sufficiently large that communication 
bottlenecks could be significant.}; it is thus useful to understand the communication requirements of algorithms for this problem. By adding the reset transitions to a large enough set $S$, it is possible to show that the harmonic walk on $G'$ is {\em rapidly-mixing}, and hence any iterative algorithm for computing the pageranks will converge quickly, typically in $O(\log n)$ iterations. As the results in this paper are primarily worst-case results, they are unlikely to be tight for this particular problem. However, we feel that it is important to understand the worst-case complexity of the more general problem as we define it.

The desirable properties of Markov chains (for example, based on the known stability of principal eigenvectors under small perturbations) have led to them finding new applications in distributed web searching \cite{wang04vldb,hpdc03pagerank}, distributed `reputation' systems \cite{eigentrust} and many other problems that can be expressed as finding the stationary distribution of a Markov chain on some network. Although several distributed algorithms have been proposed for pagerank \cite{eigentrust,hpdc03pagerank,kamvar03exploiting,page98pagerank,wang04vldb,shi03icpp}, to the best of our knowledge nothing nontrivial is known about the communication complexity of the problem, nor these algorithms.

In trying to model more faithfully browsing behaviour on the world-wide web, Fagin et al. \cite{fagin00} introduce backoff processes. Such a process can be defined by a graph $G$ (and its harmonic random walk), and for each node, a {\em backoff probability} $\rho$, where at each time step with probability $1-\rho$, the token moves forward as defined by the walk, and with probability $\rho$ it returns to its previous state. They show some interesting phenomena that are induced by this process, for example it does not always have a limit distribution independent of the starting state, even if the underlying chain is ergodic. It would seem interesting to extend our results to obtain lower bounds for the complexity of these processes.

Fogaras and Rasz \cite{fogaras04} consider a related problem known as `personalized pagerank'. The personalized pagerank for a node $u$ is defined as the unique stationary vector $$\pi(u) = (1-c) P \pi(u) + c \mathbf{u}$$ where $\mathbf{u}_u=1, \mathbf{u}_v=0$ for $v \neq u$. They prove simple lower bounds on the size of the database required for a centralized server to be able to answer queries about $\pi(u)$ (for all nodes $u$), in both exact and approximate models. Like our results, their lower bounds utilise communication complexity arguments, but they only require reductions from one-way communication complexity rather than the two-way results we require to lower bound arbitrary distributed computations. Because of this, their results are purely information-theoretic, yet for some of our problems we are able to give much stronger results than the information-theoretic bound, by showing how we can use the network to `do work' for us.

As far as we know, we are the first to consider communication complexity lower bounds for problems where the network structure itself forms the input, and the output depends nontrivially on its structure. There has, of course, been progress with proving lower bounds for problems in distributed computation. Abelson \cite{322200} obtained the first nontrivial lower bounds for a distributed protocol to solve a system of linear equations, although his result applies to differentiable real-valued functions. The lower bound is based on showing that the matrix that describes the system has sufficiently high rank that any protocol must make a large number of choices to locate the solution. This is related to the well-known `fooling set' lower bound technique now a staple part of communication complexity. However, Abelson's result gives a lower bound on the number of values that must be communicated whereas we give information-theoretic results on the number of bits that must be communicated over a network which also forms part of the input.

\subsection{Summary of our results}

In section \ref{section:reversible_chains}, we consider the case where the network $G$ is undirected and unweighted. The harmonic random walk induced is then that of a reversible chain. For these graphs we show that there is a simple optimal algorithm to
compute the stationary distribution. This result shows that reversible
chains can only encode local information about the graph into the
stationary probabilities.

Next we consider the case where $G$ is directed. Our results show that,
unlike the undirected case, interesting structural properties can be encoded into the stationary distribution, and
understanding the nature of this is our main tool in obtaining
good communication complexity lower bounds. In Section
\ref{section:general_unweighted} we prove a lower bound on the total
communication required for any distributed algorithm to compute the
stationary probability of a single node in the graph. The motivation for this result is that, in a large distributed network, it is not efficient to compute the values for all nodes if only one node requires its value. As we show, even though the stationary probability of a single node depends on the stationary probability of all other nodes, our results suggest that we can compute a single probability from scratch, at a lower cost than computing all the values.

We then consider variants of the basic problem. In
Section~\ref{section:distribution}, we look at computing
the entire stationary distribution and prove that, asympototically, there are the same number of distinct principal eigenvectors as there are unweighted graphs. In Section \ref{section:approximation} we turn to the problem of
approximately computing stationary probabilities. Currently we know of
no distributed approximation algorithm that achieves a specified
approximation factor, yet this appears to be a useful practical problem.
In Section \ref{section:computing_ranks} we consider an interesting
variant of the problem: computing the rank of a single node in the
stationary distribution $\pi$. We prove a communication lower bound for
computing the node with maximal rank, and whether a node has even or odd
rank (which implies a bound on computing the actual rank).

Finally, in comparing our results to those for computing shortest path lengths, we use the elegant path algebra framework of Gondran and Minoux \cite{817} to formalize the problems, and discuss the complexity results in terms of algebraic properties of the problem. This appears to be a novel approach to accounting for the complexity differences.

\section{Bounds for reversible chains}
\label{section:reversible_chains}
A Markov chain is \emph{reversible} iff it satisfies the detailed
balance
equations $$\pi(u) p_{uv} = \pi(v) p_{vu}$$ i.e. the probability flux between any two nodes
is the same in both directions. In particular, if $G$ is undirected then
the harmonic random walk on $G$ is a reversible Markov chain. Reversible chains have many remarkable properties. In particular, the key to efficient computation on reversible chains is the following: the stationary probability of any node is proportional to its degree -- regardless of the structure of $G$. More precisely, if $G$ is undirected then using the detailed balance equations, it is easy to verify that $\pi(u) = \mbox{degree}(u) / 2 |E|$ is indeed a stationary probability, and if $G$ is ergodic then this is unique. Hence the stationary probability $\pi(u)$ is determined solely by $|E|$ and the degree of $u$. If $G$ is not reversible then in general there is no such local expression for computing $\pi(u)$.

A simple algorithm for computing $\pi(u)$ is then as follows: given a spanning tree $T$ of $G$, each node computes the sum of the degrees of
the nodes below it in the tree in a depth-first manner. There are $n-1$ edges in $T$ and each edge carries at most $O(\log n^2)=O(\log n)$ bits, hence this algorithm sends $O(n \log n)$ bits in total in the worst case. The following simple theorem shows that this is asymptotically optimal.

\begin{theorem}
Any distributed algorithm that computes $\pi$ for an $n$-node reversible chain must communicate $\Omega(\log n)$ bits over $\Omega(n)$ edges in the worst case.
\end{theorem}
\begin{proof}
We will show an information-theoretic bound on the communication required between two sides of a cut in a sufficiently large class of graphs. Consider an undirected graph $G$ with $n$ nodes $u_1,\ldots,u_n$, and add an extra node $v$ with an edge $(u_n,v)$. We will consider the amount of communication required between $u_n$ and $v$, i.e. across the edge $(u_n,v)$.

Since the degree of $v$ is fixed and the harmonic walk on $G$ gives a reversible chain, only the number of edges between the $u_i$ affects the stationary probability of $v$. Since there are $\Omega(\binom{n}{2}-n)=\Omega(n^2)$ strongly-connected graphs with a distinct number of edges, this gives $\Omega(n^2)$ different possible values of $\pi(v)$. If each one is equally likely then at least $\Omega(\log (n^2))$ bits must cross the edge $(u_n,v)$.

Now imagine replacing the edge $(u_n,v)$ by a linear array of $n$ edges. As described in Section \ref{sec:technique}, we can lift our lower bound onto the linear array with an increase by a factor $kn$, for some constant $k$. Hence at least $\Omega(\log n)$ bits must flow over at least $\Omega(n)$ edges in the worst case.
\end{proof}

It seems that the requirement of reversibility precludes the existence of an interesting relationship between the structure of the graph and the stationary distribution of the harmonic walk on it.

\section{Directed Markov chains}
\label{section:general_unweighted}
In the remainder of the paper, we consider directed graphs. The Markov chain may not have a stationary distribution and certainly does not have a simple closed form, as for undirected graphs.

A Markov chain is said to be {\em irreducible} if for all $u,v$ there is a positive probability of the token reaching $u$ from $v$. Assume that the chain is irreducible. A fundamental result is that there exists a unique stationary distribution $\pi = (\pi(1),\pi(2),\ldots,\pi(n))$ with $\sum_u \pi(u) = 1$ that satisifies the balance equations $$\pi(u) = \sum_v \frac{1}{\mbox{outdegree}(v)} \pi(v).$$ We shall therefore assume that $G$ is strongly-connected and non-bipartite, as this will guarantee the existence of a stationary distribution.

Let $p_{uv}=1/{\mbox{outdegree}(u)}$ be the transition probability from state $u$ to $v$, and $p_{uv}^{(k)}$ be the probability of the token being at $v$ after exactly $k$ steps, starting from $u$. A state $u$ is recurrent if it is visited infinitely often in an infinitely long walk, and aperiodic if $\gcd \{ k : p_{uu}^{(k)} > 0 \} = 1$. Recurrent, aperiodic states are said to be {\em ergodic}. An irreducible chain whose states are ergodic is said to be ergodic. If the chain is ergodic then in addition, the limit $\lim_{k \to \infty} p_{ij}^{(k)} = \pi(j)$ exists and is independent of $i$. This forms the basis for iterative algorithms that compute $\pi$. Since we want to lower bound the communication of {\em any} algorithm that computes $\pi$, we shall not require ergodicity, but only that the stationary distribution $\pi$ exists.

\subsection{Lower bound}
We now show an information-theoretic lower bound on the communication complexity of computing the stationary probability of a single node. Assume that the graph $G=(V,E)$ is unweighted (hence $w_{uv}=1$ iff $(u,v) \in E$ and $0$ otherwise) and directed, and has $n$ nodes.
\begin{theorem}
Any distributed algorithm, randomized or deterministic, that computes $\pi(u)$
for some $u$ must communicate at least $\Omega(n \log n)$ bits over $\Omega(n)$
edges in the worst case.
\label{thm:main}
\end{theorem}
We will first prove an $\Omega(n)$ bound for sparse graphs then show
how it can be improved to $\Omega(n \log n)$ in the case of dense graphs.

The following two lemmas establish lower bounds on the communication complexity of a two-party version of the problem by reduction from set-disjointness. We show that, without any communication Alice and Bob can construct a graph $G$ (where the edges are partitioned between Alice and Bob), such that knowing the stationary probability $\pi(u)$ for some node $u$ allows them to solve disjointness.

\begin{lemma}
The randomized communication complexity is $\Omega(n)$ bits in the case of sparse graphs.
\label{lemma:sparse_graphs}
\end{lemma}
\begin{proof}
Given two $n$-element sets $P,Q \subseteq \{1,\ldots,n\}$, Alice and Bob construct a sparse graph $G$ as follows. Because the stationary value of $u$ will reveal the entire set $P$, Bob need not encode anything, and so his part of the graph is constant.

The graph contains $n$ nodes $x_1,x_2,\ldots,x_n$
on a cycle $x_1 \to x_2 \to \cdots \to x_n \to x_1$, and nodes $u$ and $u'$.
There is a sink node $x_a$, with edges $x_a \to x_1$ and $\{u,u'\} \to x_a$.
Finally, add two nodes $v$ and $v'$ with edges $u \leftrightarrow v$ and $u'
\leftrightarrow v'$. The cut of the graph shall be $\left<G,G-\{v,v'\}\right>$.
For each element $j \in P$, add an edge $x_j \to u$ and for each element $j
\not\in P$, add an edge $x_j \to u'$. 

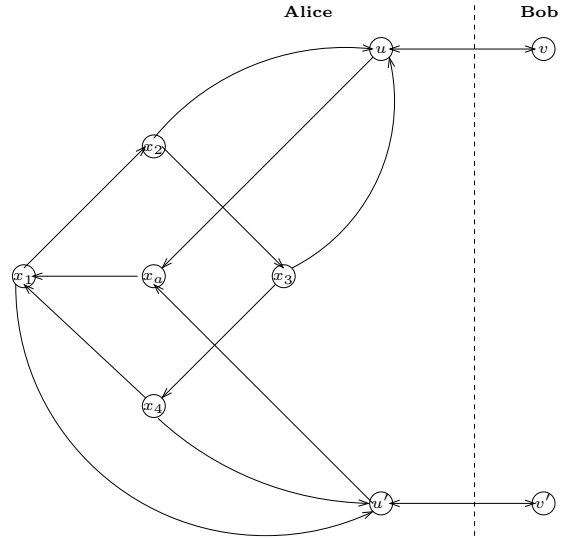
\begin{figure}[h]
\begin{center}
\input{cycle1.pstex_t}
\end{center}
\caption{Construction for Lemma~\ref{lemma:sparse_graphs} when $n=4$ and $P = \{2,3\}$.}
\label{fig:cycle1}
\end{figure}

The construction is illustrated in Figure~\ref{fig:cycle1}.
Intuitively, as the stationary probabilities halve on each step in the cycle, the binary expansion of $\pi(v)$ should reveal $P$. We shall use $\pi(j)$ to denote the stationary probability of $x_j$, and similarly $\pi(a)$ for $x_a$. Now, the flux drops exponentially along the cycle so we have
\begin{eqnarray*}
\pi(j) &=& \frac{1}{2} \pi({j-1}) \\
&=& 2^{n-j} \pi(n)
\end{eqnarray*}
Consider the stationary probability $\pi(v)$:
\begin{eqnarray*}
\pi(v) &=& \frac{1}{2} \pi(u) \\
&=& \frac{1}{2} \left [\frac{1}{2} \sum_{j \in P} \pi(j) + \pi(v) \right] \\
\Rightarrow \; \pi(v) &=& \frac{1}{2} \sum_{j \in P} \pi(j) \\
&=& \pi(n) \sum_{j \in P} 2^{n-j-1}
\end{eqnarray*}

For node $v$ to be able to obtain $P$ from the binary expansion of $\pi(v)$, the
value $\pi(n)$ must be a constant, independent of the sets $P$. Note that this
does not follow trivially; for example, if we replaced the edge between $u',v'$, with a self-loop at $u'$ (or even with nothing), then $\pi(n)$ would depend on
$P$. The reason for this is that there must be a flow of constant flux
(independent of $S$) from the nodes on the cycle, and back to $x_a$, and hence
$\pi(u) + \pi({u'})$ must also be a constant. We now show that for our
construction, this is indeed the case. For simplicity, let $s=\pi(u) + \pi({u'})$
and $t=\pi(v) + \pi({v'})$. Then
\begin{eqnarray*}
s/2 = \pi(a) &=& 1/2 ( \pi(u) + \pi({u'}) ) \\
&=& \frac{1}{4} \pi(n) \sum_{j=1}^n 2^{n-j} + \frac{1}{2} t \\
&=& \frac{1}{4} \pi(n) (2^n - 1) + \frac{1}{2} t
\end{eqnarray*}
and
\begin{eqnarray*}
t = \frac{1}{2} (\pi(u) + \pi({u'})) &=& \frac{1}{2} s \\
&=& \frac{1}{4} (2^n - 1) \pi(n) + \frac{1}{2} t \\
\Rightarrow \; t &=& \frac{1}{2} (2^n - 1) \pi(n)
\end{eqnarray*}
Since $\pi$ is a probability distribution, it must sum to unity:
\begin{eqnarray*}
\pi(a) + s + t + \pi(n) (2^n - 1) &=& 1 \\
\frac{3}{2} s + t + \pi(n) (2^n - 1) &=& 1 \\
\frac{7}{4} (2^n - 1) \pi(n) + \frac{5}{2} t &=& 1 \\
3 (2^n - 1) \pi(n) &=& 1 \\
\pi(n) &=& \frac{1}{3 (2^n - 1)} = c
\end{eqnarray*}
Hence $\pi(v) = c \sum_{j \in P} 2^{n-j-1}$ for some constant $c$. Now, suppose
node $v$ computed $\pi(v)$ in this graph. We can assume that it knows $c$ (as it
depends only on $n$). Then it can read off the $n$-bit set $P$ from the binary
expansion of $\pi(v)$ (note that the largest element of $P$ is represented by the least significant bit of $\pi(v)$). Since the randomized communication complexity of set-disjointness is $\Omega(n)$ bits, at least this many bits must cross the cut between Alice and Bob.
\end{proof}
More precisely, the construction defines a class of sparse graphs where any algorithm that computes $\pi(v)$ with probability of error at most $p$ allows some node to learn an $n$-bit set with the same probability of error.

Now we show how the two-party lower bound can be improved to $\Omega(n \log n)$ bits in the case of dense graphs. The idea is that, instead of each node on the cycle encoding a single bit, each node can encode a small set of size $O(\log n)$ bits, since it can potentially transfer $O(n)$ different proportions of its flux to the node $u$. These small sets are now encoded in blocks of $O(\log n)$ bits into $\pi(u)$.

\begin{lemma}
\label{lemma:denselowerbound}
The randomized communication complexity is $\Omega(n \log n)$ bits, for dense graphs.
\end{lemma}
\begin{proof}
Take the previous construction, and add $2n$ nodes $z_1,z'_1,\ldots,z_n,z'_n$,
where each $z_i$ links to $u$ and each $z'_i$ links to $u'$. Now partition the
set $P$ into $\log n$-element sets $P_1,\ldots,P_n$ where $P_j$ will be encoded
by node $x_j$. Each node $x_j$ on the cycle links to exactly $n+1$ nodes: It
links to $x_{j+1}$, and for
each $i$, it links to either $z_i$ or $z'_i$.

Since the edges are unweighted, each edge contributes the same flux from a
node, hence each node $x_j$ can now give $n$ different fractions of its
probability flux to $u$, via the $z_i$. The intuition is that each $x_j$ can
independently encode a set of $O(\log n)$ elements, allowing us to encode $O(n
\log n)$ bits of information into $\pi(v)$. As before, we also need to show that
$\pi(n)$ is still constant.

The flux now drops by a factor $n+1$ on each step of the cycle, so
$\pi(j)=\frac{1}{n+1} \pi({j-1}) = (n+1)^{n-j} \pi(n)$. Hence $\sum_{j=1}^n \pi(j)= \pi(n) ((n+1)^n - 1)$. Let $d(P_j)$ denote the value of the binary
expansion of the set $P_j$ (where $0 \leq d(P_j) \leq 2^{|P_j|}$). Then each
node $x_j$ links to exactly $d(P_j)$ of the $z_i$ (and hence exactly $n-d(P_j)$
of the $z'_i$). Consider the stationary probability $\pi(v)$:
\begin{eqnarray}
\pi(v) &=& \frac{1}{2} \pi(u) \nonumber \\
&=& \frac{1}{2(n+1)} \sum_{j=1}^n \pi(j) d(P_j) + \frac{1}{2} \pi(v) \nonumber \\
&=& \frac{1}{n+1} \sum_{j=1}^n \pi(j) d(P_j) \nonumber \\
\label{eqn:dense_expansion}
&=& \frac{\pi(n)}{n+1} \sum_{j=1}^n (n+1)^{n-j} d(P_j)
\end{eqnarray}

Now we show that the flux crossing the cut is indeed constant. As before, let $s=\pi(u) + \pi({u'})$ and $t=\pi(v) + \pi({v'})$. Then
\begin{eqnarray*}
\frac{1}{2} s = \pi(a) &=& \frac{1}{2} ( \pi(u) + \pi({u'}) ) \\
&=& \frac{n}{2(n+1)} \pi(n) ((n+1)^n - 1) + \frac{1}{2} t
\end{eqnarray*}
and
\begin{eqnarray*}
t = \frac{1}{2} (\pi(u) + \pi({u'})) &=& \frac{1}{2} s \\
&=& \frac{n}{n+1} \pi(n) ((n+1)^n - 1) \\
\end{eqnarray*}

Since $\pi$ is a probability distribution, it must sum to unity:
\begin{eqnarray*}
\frac{3}{2} s + t + \pi(n) ((n+1)^n - 1) &=& 1 \\
\frac{5}{2} \frac{n}{n+1} ((n+1)^n - 1) \pi(n) + ((n+1)^n - 1) \pi(n) &=& 1 \\
((n+1)^n - 1) (\frac{5}{2} \frac{n}{n+1} + 1) \pi(n) &=& 1 \\
\end{eqnarray*}
and so $\pi(n)$ is a constant $c$ independent of $P$. Hence $\pi(v) = c
\sum_{j=1}^n (n+1)^{n-j-1} d(P_j)$. Since the sets $P_j$ are of $\Theta(\log
n)$ bits, $d(P_j) \leq n$ and hence $\pi(v)$ reveals all the values $d(P_j)$,
and hence the $n$ sets $P_j$.
\end{proof}

The previous lemmas have given us lower bounds for the transfer across the cut between two parties, where Alice and Bob each know only their subgraph. Note that we have taken care that in both constructions, the cut $\left< X,Y \right>$ is sparse and so we can build a modified $O(n)$-node graph by replacing each edge in this cut by $n$ edges.

We now appeal to the linear array conjecture and so in the worst case at least $\Omega(n \log n)$ bits must flow over each of these edges (using the constant in the lifting of randomized communication bounds onto a linear array).

\subsubsection{Remarks}
The expansion of $\pi(v)$ represented by Equation \ref{eqn:dense_expansion} gives a clue as to why we cannot hope to improve the lower bound using our current methods. See that it can be roughly rewritten as (ignoring constants) $\sum_{j=1}^n \left(\frac{1}{2}\right)^{j \lg n} d(P_j)$. Hence each $P_j$ only has $\approx \lg n$ bits in the expansion of $\pi(v)$, even though it is quite easy to build a construction where the $P_j$'s can be of $n$ bits, and in this case the sets cannot be recovered since they begin to overlap in the binary expansion of $\pi(v)$.

The result also gives lower bounds on the worst-case congestion and time incurred by any algorithm to compute $\pi(v)$.For congestion, the stretching trick means that there must be at least a linear number of edges that each have $\Omega(n \log n)$ bits communicated across them. For a time lower bound, there are $\Omega(n \log n)$ bits that must cross a cut of size $O(1)$, hence any algorithm must take time $\Omega(n \log n)$ in the worst case.

\subsubsection{Labelling schemes}

A simple and interesting corollary of the two-party lower bound in this section is a lower bound on the complexity of a labelling scheme. A {\em distance labelling scheme} for a graph $G$ is an assignment of labels $l(v)$ to nodes $v$ of $G$ such that, by examining only the labels $l(u),l(v)$, one can determine the distance $d(u,v)$ between $u$ and $v$. By encoding global information about a graph into local labels, labelling schemes have many practical applications in large-scale distributed networks \cite{gavoille01distance}. A {\em hitting time labelling scheme} is the natural extension of a distance labelling scheme to the random walk on a graph: given $l(u),l(v)$, one can compute the expected length $E[d(u,v)]$ of the random walk beginning at $u$ and terminating on first hitting $v$.

Since $\pi(u) = 1 / E[d(u,u)]$, Lemma \ref{lemma:denselowerbound} implies that for any hitting time labelling scheme, there must be some graph where some node must be assigned a label of size $\Omega(n \log n)$ bits (in particular, this must occur for some node computing the expected time for the token to return to itself). Clearly there is an upper bound of $O(n^2)$ on the size of labels (by encoding the whole graph into each label), but the aim of efficient labelling schemes is to do much better than this. In \cite{gavoille01distance} it is shown that $\Theta(n)$ bits is the optimal distance label length for general unweighted graphs, so our result shows an increase in complexity, but we do not know if the increase is more than just a logarithmic factor.

\subsubsection{Upper bounds for computing $\pi$}

A simple algorithm to compute $\pi(u)$ would be to construct a spanning tree $T$ of $G$ rooted at node $u$, and for each node to send a description of its edges to $u$ using $T$. For general unweighted graphs, this would require $O(n^2)$ bits being sent over $O(n)$ edges in total. Constructing a distributed algorithm with $o(n^3)$ bits worst-case total communication appears to be a challenging problem. We conjecture that the true lower bound is $O(n^2 \mbox{polylog}(n))$ but have been unable to prove this for an algorithm. We also believe that, for the problem of {\em exactly} computing the stationary probabilities, randomization is of no help as regards worst-case communication complexity.

For the related shortest paths problem, there is a long and interesting history of efficient algorithms, both sequential and more recently, distributed. Understanding these may help in obtained nontrivial upper bounds for the path problems we consider here. The best known communication complexity upper bound in the distributed case is $O(n^2 \log^2 n)$ bits and relies on a graph decomposition to represent the graph as a partition of sparsely-connected clusters \cite{afek92sparser}. It is reasonably easy to show that any distributed algorithm that computes the shortest path lengths must have worst-case communication complexity $\Omega(n^2 \log n)$ bits (there exist graphs where the length of the path to each node requires $\Omega(\log n)$ bits, and each must be sent over $\Omega(n)$ edges). Hence its communication complexity is fairly well-understood.

\section{Computing the entire distribution}
\label{section:distribution}

In this section we consider the slightly different problem of some node $v$ knowing the entire vector $\pi$ of stationary probabilities. This may correspond
to a distributed crawling algorithm that terminates with some centralized server $v$ knowing the entire vector.

Our results for this section illustrate an interesting weakness of our
lower bound technique. In the two-party case, we prove that the trivial
algorithm of sending the entire graph is optimal (and so we cannot do
any better here), but since the cut between the $x_i$ and $y_i$ nodes is
dense, we cannot amplify our bound by lifting onto a linear array (or
other sparse structure). Therefore we only obtain an $\Omega(n^2)$ bound
in the distributed case, even though the trivial (spanning tree)
algorithm costs $O(n^3)$ in this model. Improving this situation with
our current technique would involve finding a construction with the same
two-party complexity but with a much sparser cut, say with a constant or
(poly)logarithmic number of edges. We feel that an $O(n^2)$ bound is not
possible with this number of edges.

The intuition for an information-theoretic lower bound might be something like
this: each edge in the graph can alter the vector $\pi$, therefore there are
$2^{n^2/ 2}$ possible vectors, so $\Omega(n^2)$ bits must be communicated. Of
course, this is nontrivial because while each edge does indeed change $\pi$, it
is still possible that many combinations of edges result in the same $\pi$. For
example, an $n$-clique and an $n$-cycle (adding chords to make it ergodic) both
have the same $\pi$. So what we need to prove is a bound on the number of {\em
distinct} vectors $\pi$ (or, the number of distinct principal eigenvectors of a
set of $n$-node graphs).

We prove the lower bound by exhibiting a family of $n$-node graphs with
$2^{\Omega(n^2)}$ distinct principal eigenvectors.
\begin{theorem}
There is a family of $n$-node directed, unweighted Markov chains with $2^{n^2}$ distinct stationary vectors $\pi$.
\label{thm:cycledense}
\end{theorem}
\begin{proof}
The construction is as follows. There are nodes $x_1, x_2, \ldots, x_n, y_1, y_2, \ldots, y_n$, $y'_1, y'_2, \ldots, y'_n$ and a sink node $s$. The edges are as follows:
\begin{itemize}
\item a cycle $x_1 \to x_2 \to x_3 \to x_n \to x_1$
\item $s \to x_1$, to get an exponential dropoff on the cycle
\item $y_i \to s$ and $y'_i \to s$, for all $i$
\end{itemize}
Finally, call a matrix $V=v_{ij}$ legal if $v_{1j}=1$ and $v_{2j}=0$ for all $j$. For each entry $v_{ij}$, if $v_{ij} = 1$, add the edge $x_i \to y_j$ and if $v_{ij} = 0$, add $x_i \to y'_j$. First, note that this graph is strongly connected for all legal matrices (this is why we forced the first row elements to 1, second row to 0).
The construction is illustrated in Figure~\ref{fig:cycledense}.

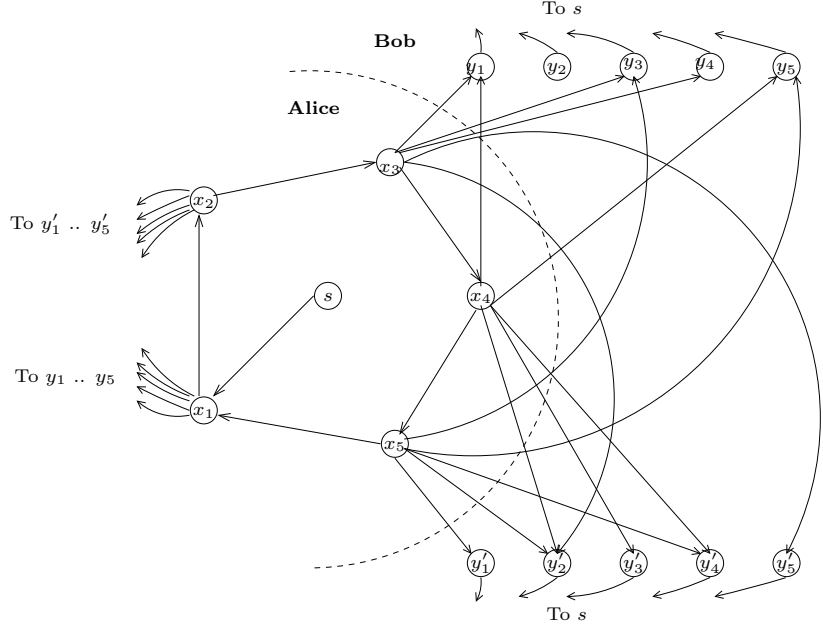
\begin{figure*}[t]
\begin{center}
\input{cycledense.pstex_t}
\end{center}
\caption{Construction for Theorem~\ref{thm:cycledense}.}
\label{fig:cycledense}
\end{figure*}

We now show that each legal matrix gives a different vector $\pi$, over the nodes $y_i$. Firstly, see that if we flip one bit $v_{ij}$, the only values that change are $\pi({y_j})$ and $\pi({y'_j})$. Now we show that each distinct vector $(v_{3j}, v_{4j}, \ldots, v_{nj})$ gives a different value for $\pi({y_j})$.

To the contrary, assume there are two vectors $v \neq v'$ with $\pi^v({y_j}) = \pi^{v'}({y_j})$, where $\pi^v({y_j})$ is the stationary probability of $y_j$ under the vector $v$ (the $j$th column of the matrix $v_{ij}$). By definition,
\begin{eqnarray*}
\pi^v({y_j})     &=& \frac{1}{n} (\pi({x_1}) + \sum_{i=3}^n v_i \pi({x_i})) \\
                &=& c + \sum_{i=1}^n v_i \pi({x_i}) ~~~\mbox{for some constant $c$} \\
                &=& c + \sum_{i=1}^n v'_i \pi({x_i}) ~~~\mbox{by assumption} \\
                &=& \pi^{v'}({y_j})
\end{eqnarray*}
that is, if $\pi^v({y_j}) = \pi^{v'}({y_j})$ then there must be two distinct subsets of $(\pi({x_1}), \pi({x_2}), \ldots, \pi({x_n}))$ that sum to the same value. But this is impossible since the values fall off exponentially (with the same factor) on the cycle construction.

This proves that each legal matrix gives a different vector $\pi$, therefore the number of different vectors $\pi$ is equal to the number of different vectors $(v_{ij})$, which is $2^{n(n-2)}$. 
\end{proof}

The above lemma shows constructively that there are a family of n-node graphs (the whole construction) with $2^{\Omega(n^2)}$ distinct stationary vectors (taken over the $O(n)$-node subgraph of the $y_j$'s) and so any node that is to know this vector must have at least $\Omega(n^2)$ bits communicated to it in the worst case.

\section{Approximate computation}
\label{section:approximation}
In this section we turn to the natural problem of approximating the stationary probabilities $\pi(v)$. Firstly, see that we must be careful with our notion of approximation: if $\pi(v)$ is to be computed to within $k$ bits of precision, we can use the construction of Lemma \ref{lemma:sparse_graphs} to encode a set of size $O(k)$ bits rather than $O(n)$ bits, and the lower bound is accordingly reduced to $\Omega(k)$ bits across $\Omega(n)$ edges. A more natural notion of approximation may be to compute $\pi(v)$ to within a given factor. Call $\hat\pi(v)$ a $k$-approximation to $\pi(v)$ if $\frac{1}{k} \pi(v) \leq \hat \pi(v) \leq k \pi(v)$. In this section we prove that any distributed algorithm that computes a $k$-approximation to $\pi(v)$ for some chosen $v$ (even with high probability) must send send at least $\Omega(\log \frac{n}{\log k})$ bits across $\Omega(n)$ edges in the worst case.

First, let us examine the case $k=2$. See that the difficulty with using our original binary encoding of the set is that, in the binary representation, all the bits of lower order than the highest order bit can be changed arbitrarily while remaining within a factor $2$. The basic idea is to use a simple error correcting code that resists changes to a numeric factor of 2. Just using the highest order bit is not sufficient, since for example $0100$ can become $1000$ (an increase by a factor 2) and $0010$ (a decrease by a factor 2). A simple solution is to pad out the expansion, using blocks of length 3 bits, for example $$\cdots00\underbrace{{\bf 010}}_{\mbox{a block}}00\cdots$$ Then the highest order bit can never fall out of its block. In the lower bound, we use blocks of length $1 + 2 \lceil \lg k \rceil$ bits to withstand factor $k$ approximations.

The idea then, is to use a variant of the construction from Lemma \ref{lemma:denselowerbound} to produce a binary string where the set $S$ is encoded into the position of the highest order bit of the `blocked' binary expansion. Just considering encoding a single set $S$, we are looking for a binary expansion of the form $$0\ldots 00{\bf 010}\underbrace{00 \ldots 0}_{d(S)}$$ where the 1 is at some position, determined by the value (as previously defined) $d(S)$ of the set. Encoding $O(\log n)$ elements requires $O(2^{\log n}) = O(n)$ possible indices in a binary expansion, which is exactly what we have available from the original construction.

To encode an $O(\log n)$ bit set $S$, we compute $d(S)=\sum_{j \in S} 2^j$ and then use the node $x_{d(S)}$ to encode this value by linking it to $u$, and all other nodes to $u'$. The claim is that the block containing the highest order bit (and hence the encoding of the set) of the binary expansion of $\pi(v)$ can be recovered from the binary expansion of $\pi^k_v$. Let us now prove the main lower bound.
\begin{theorem}
Consider any distributed algorithm computes $\hat\pi(v)$ with $\Pr(\frac{1}{k} \pi(v) \leq \hat \pi(v) \leq k \pi(v)) > 2/3$. Then it must send $\Omega(\log \frac{n}{\log k})$ bits over $\Omega(n)$ edges in the worst case.
\end{theorem}
\begin{proof}
We extend the idea outlined above to an approximability-preserving reduction from set-disjointness. The main difference is the `block cycle' construction: for each node $x_1,\ldots,x_n$ on the cycle, replace it with a block of $\kappa=(1 + 2 \lceil \lg k \rceil)$ nodes where $x_i$ is now the center node of the $i$th block (and the rest are dummy nodes). We will show how to encode a $O(\log n)$ bit set $S$ using $O(n \kappa)$ nodes in the construction. Imagine that the set has value $d(S)=j$. Now, pick the node $x_j$ on the cycle and add an edge $x_j \to u$, and for all other nodes on the cycle (including the dummy nodes) add an edge to $u'$. The block construction will let $v$ recover the value $j$ under a $k$-approximation. The ratio between successive $x_i$'s on the cycle is now $2^\kappa$, hence
\begin{eqnarray*}
p_j &=& \frac{1}{2^\kappa} p_{j-1} \\
&=& p_n (2^\kappa)^{n-j} \\
&=& p_n 2^{\kappa (n-j)}
\end{eqnarray*}

A bit of algebra establishes that $p_n$ is indeed a constant $p_n = 1/(3(2^{n \kappa} - 1))$. Consider now the actual value $\pi(v)$. Imagine encoding the set $S$. Let $j$ be such that the node $x_j$ will represent this set (as described above). Then
\begin{eqnarray*}
p_v = p_u / 2 &=& p_j / 2 + p_v / 2 \\
&=& p_j / 2 \\
&=& p_n 2^{\kappa (n-j) - 1} \\
&=& \frac{1}{3} \frac{1}{2^{\kappa j + 1}} \\
\end{eqnarray*}

Now $v$ can find $j$ easily, since $2^{\kappa j} = (2^{1 + 2 \lceil \lg k \rceil})^j = (2k^2)^j$, and $\log_{2k^2} (2k^2)^j = j$. Now we claim that the block containing the highest order bit is the same in both $\pi^k_v$ and $\pi(v)$, by showing that the values must be separated by a factor of least $2 k^2$. Assume the set being encoded has value $j$, hence (ignoring constants) the largest $k$-approximate value of $p_v$ is
\begin{eqnarray*}
k p_v &=& \frac{1}{2^j} \frac{k}{k^{2j}} \\
&=& \frac{1}{2^j} \frac{1}{k^{2j-1}}
\end{eqnarray*}
Now if the set had value $j-1$ instead, then the smallest $k$-approximate value would be (letting $\pi'_v$ be $v$'s stationary probability using the set with value $j-1$)
\begin{eqnarray*}
p'_v / k &=& \frac{1}{2^{j-1}} \frac{1}{k^{2(j-1)+1}} \\
&=& 2 \frac{1}{2^j} \frac{1}{k^{2j-1}} \\
&=& 2 k p_v
\end{eqnarray*}

These two values are the closest possible (a $k$-overapproximation with set $j$ and a $k$-underapproximation with set $j-1$) and are still separated by exactly one bit in their binary expansion, and so there is no overlap and the value of the set can be recovered. Intuitively, the binary expansions of $\pi^k_v$ and $\pi(v)$ look like the following, for a set $S$ with value $d(S)=j$.
\begin{eqnarray*}
\pi^k_v && \cdots 000\underbrace{01101}_{\mathrm{block}~ j}\underbrace{01011}_{j-1}010010 \cdots \\
\\
\pi(v) && \cdots 000\underbrace{00100}_{\mathrm{block}~ j}00000000000 \cdots \\
\end{eqnarray*}
where each block has $\kappa$ bits and the highest order bit of $\pi^k_v$ is contained in block $j$ iff the highest order bit of $\pi(v)$ is in block $j$.

For the communication complexity bound, see that there are only a constant number of edges crossing the cut, and so any protocol that computes a $k$-approximation $\hat\pi$ with probability $p$ allows us to solve $O(\log n)$ disjointness with $O(n \kappa)=O(n \log k)$ nodes, with the same probability. Hence, for a graph of $n$ nodes we can solve $\Omega(\log (n/\log k))=\Omega(\log n - \log \log k)$ disjointness. The result follows since the cut is sparse and we can appeal to the linear array result, and by the communication complexity of set-disjointness.
\end{proof}

\subsubsection{Remarks}
As before, the result yields analogous time and congestion lower bounds. Also, note that $\Omega(\log n - \log \log k)=\Omega(\log n)$ for constant $k$. It may be interesting to investigate what happens for $k=1+\epsilon$.

\section{Computing the ranks}
\label{section:computing_ranks}
We say that a node $u$ has rank($u$)=$k$ iff there are exactly $k$ nodes $\{v_1,\ldots,v_k\}$ with stationary probabilities at least as large as $u$: $\pi(u) \leq \pi({v_i})$ for $i=1 \ldots k$. Hence $u$ has maximal rank iff rank($u$)=1, and it has minimal rank if rank($u$) = $n$.

In this section we consider the difficulty of computing the rank of some node. Clearly if there are $n$ nodes then the rank of a node can be expressed with $O(\log n)$ bits, unlike the stationary probability, which by Lemma \ref{lemma:denselowerbound} can require $\Omega(n \log n)$ bits. On the other hand, knowing that rank$(u)=k$ implies that some node must know there are exactly $k-1$ nodes having larger stationary probability and $n-k+1$ nodes having smaller stationary probability.


We now investigate the case where some algorithm terminates with many nodes in the network knowing their ranks -- again we prove a lower bound via a two-party argument and a lifting of the lower bound onto a linear array. In the two-party case, both Alice and Bob will need to know the ranks of $\Omega(n)$ nodes in each of their subgraphs.

In fact, the lower bound holds when only the parity of the ranks are known, i.e. whether a node's rank is odd or even. The lower bound shows
that at least $\Omega(n^2)$ bits must be communicated in total. Although our original intention was to prove a result for knowing the exact rank, we have been unable to strengthen it beyond the result here.

Given this statement, a natural question is `what use is knowing whether your rank is odd or even, since it doesn't imply anything
about knowing how many nodes have larger or smaller stationary value than you
(except for the parity of this number)?' We feel that presenting the result in this form exposes more details about the problem, and our proof.

On the other hand, if each node
knows whether it has even or odd rank, this may provide a useful
partitioning or symmetry breaking of the network into two pieces where the
total stationary probability in each piece is approximately equal (since if
one side has the maximum node then the next largest will be in the other
side). An interesting thing would be to determine if this can be done
without explicitly computing the ranks first.

The following theorem shows that the communication complexity of computing the rank parities is surprisingly large.

\begin{theorem}
Consider any algorithm that terminates with each node $v_i$ knowing $\mbox{rank}(v_i) \mod 2$. The communication complexity is $\Omega(n)$, and this many bits must be sent over $\Omega(n)$ edges in the worst case.
\label{thm:rank}
\end{theorem}
\begin{proof}
To construct the two-party problem, partition the network into $\left<X,Y\right>$ where Alice knows $X$ and Bob knows $Y$, and form an `exponential cycle' in $X$ with nodes $x_0 \to x_1 \to \cdots \to x_{2n+1}
\to x_0$, and add an edge from the sink $s \to x_0$. For each $i$, add
edges $x_0 \to a'_i$, $x_{2n+1} \to a_i$ and $a_i \to s, a'_i \to s$. Now,
for each $i$, if $i \in P$, add edges $x_{2i} \to a_i, x_{2i+1} \to a'_i$
else add edges $x_{2i} \to a'_i, x_{2i+1} \to a_i$. The point of the
construction so far is that all the $a'$ nodes have higher stationary
probability than all the $a$ nodes. The $Y$ partition is exactly symmetric,
with $y_{2i} \to b_i, y_{2i+1} \to b'_i$ if $i \in Q$ else $y_{2i} \to
b'_i, y_{2i+1} \to b_i$. Finally, connect the two partitions with a sparse
cut by adding edges $s \leftrightarrow t$ between the two sinks.

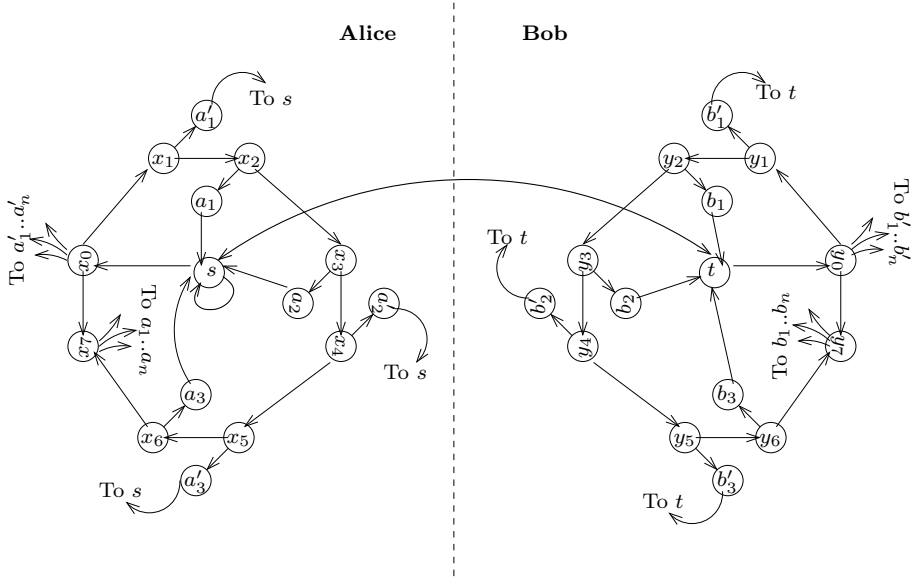
\begin{figure*}[t]
\begin{center}
\input{rank.pstex_t}
\end{center}
\caption{Construction for lower bound on computing ranks
(Theorem~\ref{thm:rank}).}
\label{fig:rank}
\end{figure*}

So far, the construction is completely symmetric. But we want the $b_i$
nodes in $Y$ to have slightly lower stationary probabilities, so we add a
self-loop at node $s$. The construction is illustrated in
Figure~\ref{fig:rank}.
Now, we consider the rankings of each node in the
construction. There are three claims:
\begin{enumerate}
\item The rankings of the $x_i, y_i, s, z, t$ are independent of $P,Q$. This follows since the stationary probabilities of these nodes are constant (in the same way as for the construction of Lemma \ref{lemma:sparse_graphs}), and for each $i$, rank($x_{2i}$) and rank($x_{2i+1}$) are always the same apart, i.e they are always separated by the $a_i,a'_i$ nodes, and the $y_i$ nodes (which also have constant stationary probability). Finally, adding the self-loop at $s$ only changes the relative stationary probabilities of the two sides, since both $\pi(s),\pi(t)$ are constant.
\item rank($a'_i$) $<$ rank($a_j$) and rank($b'_i$) $<$ rank($b_j$), for all $i,j$. This is because the $a'_i$ and $b'_i$ tap into the $x_0$ and $y_0$, which have the highest stationary probability on the two cycles.
\item For any sets $P,Q$, rank($a_i$) is less than both rank($a_{i+1}$) and rank($b_{i+1}$), i.e $\pi({a_i})$ is larger. Also, rank($b_i$) is less than both rank($a_{i+1}$) and rank($b_{i+1}$). This is because the probability flux drops off exponentially along the cycle.
\end{enumerate}

So, the only effect of the sets $P,Q$ on the rankings of the $a_i$ and $b_i$ is to change the relative rankings of $a_i$ and $b_i$, for each $i$. 
\begin{lemma}
If $i \in P$ and $i \in Q$ then rank($a_i$) $<$ rank($b_i$). On the other hand, if $i \not \in P, i \in Q$ then rank($a_i$) $>$ rank($b_i$).
\end{lemma}
\begin{proof}
The first part follows from the fact that the $a_i$ nodes in Alice's half (encoding $P$) have higher stationary values than the $b_i$ nodes in Bob's half (due to the self-loop at $s$). For the second part, we need that the ratio $\pi({x_i})/\pi({y_i}) < 2$. This is true because $p_s = 2/3 p_s + p_t = 3/2 p_t$ by the self-loop at $s$ and $\pi({x_i}) / \pi({y_i}) = (\pi(s) / 3^i)(3^i / \pi(t))= 3/2$.
\end{proof}

Hence, for any $i$ such that $i \in Q$, node $b_i$ can inspect whether its ranking is odd or even, and accordingly determine whether $i \in P$. From this, Alice can determine the set $Q$ held by Bob, and then can determine the inner product of $P$ and $Q$. Since we have managed to ensure that the cut is sparse, we can lift our result onto a linear array of size $n$ and the lower bound follows.
\end{proof}

Interestingly, in the construction, it is easy for each $a_i$ node to compute its stationary probability, since it only depends on the structure of the $X$ partition (and hence can be computed with no communication between the two partitions), but to compute the rank of $a_i$ is difficult because it depends on the structure of the other partition and there is an interplay between the two sides of the network. So, although the nodes do not have to compute the stationary probabilities, they must still implicitly know something about the structure of the ordering of the stationary probabilities in the network.

\subsubsection{Remarks}
It may be possible to improve the bound by encoding information into the $n!$ orderings of the ranks of the $a_i$ nodes (which would intuitively allow us to solve the greater-than problem on $\Omega(n \log n)$ bit sets), but it appears a challenging problem to achieve this with only a sparse cut. Without a sparse cut, we would be unable to appeal to the linear array conjecture to lift the bound onto a network.

\subsection{Computing the maximum node}

In this section we consider the problem of computing the node of maximal rank, i.e. for some $v$, an algorithm that terminates with at least one node $u$ knowing if $v$ is of maximum rank.


The node with maximal rank in a Markov chain is analogous to the center of a network in the shortest paths framework\footnote{The center of a graph $G$ is the set of nodes of maximal graph eccentricity, and the eccentricity of a node $v$ is $\max_u d(u,v)$.}. There appear to be several interesting applications for algorithms for finding the maximal rank node in a chain. For example, in a distributed network one could select a node with maximal rank to store a file, or to act as a leader of a subset of nodes\footnote{For reversible chains the stationary probability is proportional to the degree of a node so computing the node of maximal rank is equivalent to traditional leader election -- elect the leader as the node with highest degree, and of course the degree is known to each node.}.

\subsubsection{Lower bound}
We will prove that any deterministic distributed algorithm must send $\Omega(n^2 \log n)$ bits in total, and must cause congestion of $\Omega(n \log n)$ on at least $\Omega(n)$ links. In the deterministic case, this is as strong as the lower bound of Theorem \ref{thm:main} for exactly computing the stationary probability at $v$ (although we have been unable to show that deciding if $v$ is of maximal rank is at least as hard as computing $\pi(v)$).

The output size of the problem is a single bit, so a simple information-theoretic bound would be far from strong. However, we can do much better by showing how the network can do some useful computation for us. We first prove an easy lower bound on the complexity of the following problem. There are two nodes $u,v \in G$, and some node (in particular, this could be one of $u,v$) wants to decide whether $\pi(u) \geq \pi(v)$.
\begin{lemma}
\label{lemma:max}
Any algorithm that terminates with some node knowing whether $\pi(u) \geq \pi(v)$ for two nodes $u,v$ can solve greater-than on numbers of size $O(n \log n)$-bits, in the worst-case.
\end{lemma}
\begin{proof}
We shall use a modified version of the construction used in Lemma \ref{lemma:denselowerbound}, except that Bob's side will encode a set $Q$, and Alice will be able to determine which set is greater by looking at the value of $\pi(v)$. Let Bob build his side in a symmetric manner to Alice, using his set $Q$. Now, on both sides, only the even nodes on the cycle are used to encode the elements (as opposed using all the nodes as in the construction of Lemma \ref{lemma:denselowerbound}).


These sets are encoded in reverse, so the least signficant bit of the stationary probability represents the largest element of the set. More precisely, we can show (as in the proof of next theorem) that in the simplified case (ignoring constants) that the difference of the stationary values elegantly encodes the difference of the sets:
\begin{eqnarray*}
\pi(u) - \pi(v) &=& c' \left( 3 \sum_{j \in P} 2^{-2j} + \sum_{k \in Q} 2^{-2k} \right) \\
&& {} - c' \left( 3 \sum_{k \in Q} 2^{-2k} + \sum_{j \in P} 2^{-2j} \right) \\
&=& 2 c' \left( \sum_{j \in P} 2^{-2j} - \sum_{k \in Q} 2^{-2k} \right)
\end{eqnarray*}
Since the elements are encoded in reverse, we define the `reversed' set $\hat P = \{j \in P ~|~ (|P|-j+1) \in P\}$. Now suppose some node knows which of $u,v$ has highest rank:
\begin{eqnarray*}
\pi(u) - \pi(v) > 0 &\iff& \sum_{j \in P} 2^{-2j} > \sum_{k \in Q} 2^{-2k} \\
        &\iff& \sum_{j \in P} 2^{-j} > \sum_{k \in Q} 2^{-k} \\
        &\iff& \hat P > \hat Q
\end{eqnarray*}
The difference $\pi(u) - \pi(v)$ then reveals the difference $\hat P - \hat Q$, where $\hat P,\hat Q$ are taken wrt their binary expansions. Since there is a bijection between $P$ and $\hat P$ there is no loss in using this representation.

Therefore if Alice knows whether $\pi(u) \geq \pi(v)$, she can solve greater-than on numbers of $O(n \log n)$ bits (and with the same probability and error if they use a randomized protocol).
\end{proof}
Since the randomized communication complexity of greater-than is $\Theta(\log n)$ yet any deterministic protocol must communicate at least $\Omega(n)$ bits, the lemma suggests that randomization may be of some help in solving this problem. As before, we will appeal to the linear array result to lift our two-party lower bounds onto linear arrays to obtain bounds for the distributed case.

If there are no ties, then the lemma immediately implies the same communication bound for computing whether rank($u$) $>$ rank($v$). Now we can use the lemma to show the same lower bound for some node determining which side of a (specified) partition the node of maximal rank lies in. The idea is simple but the algebra tedious -- if we can modify the construction to force the top two nodes $u,v$ to lie in opposite sides of the partition, then knowing which side contains the maximum implies knowing whether $\pi(u) > \pi(v)$ or not, and the same result as in the lemma applies.
\begin{theorem}
\label{thm:ranks}
Consider a partition $\left<X,Y\right>$ of a graph. Any deterministic (randomized) distributed algorithm that terminates with at least one node knowing whether the node of maximal rank is in $X$ must communicate at least $\Omega(n \log n)$ bits ($\Omega(\log n + \log \log n)$ bits) over $\Omega(n)$ edges in the worst case. In particular, this applies if at least one node $v$ knows if it has maximal rank.
\end{theorem}
\begin{proof}

\begin{figure*}[t]
\begin{center}
\input{ranks.pstex_t}
\end{center}
\caption{Construction for Theorem~\ref{thm:ranks}.}
\label{fig:ranks}
\end{figure*}

The construction is based on the construction of the previous lemma, except that we want $v$ to be able to determine the answer to an instance of greater-than by knowing if $v$ has maximal rank. An interesting feature of the construction is that it does more than just encoding a set into the result; the network itself actually does some computation in solving the greater-than instance, and the result appears at $v$. To achieve this, we need to build the construction twice, once in each partition of the network. This is quite a powerful idea and allows us to substantially improve on the purely information-theoretic lower bounds.

Let $P,Q \subseteq \{1,\ldots,n\}$, and all nodes in $X$ (Alice's subgraph) know only $P$ and all nodes in $Y$ (Bob's subgraph) know only $Q$. The idea is that if the node with maximal rank is in the $X$ partition then $P>Q$ (wrt their binary expansions), otherwise $P<Q$. We shall use the construction of the previous lemma, but with a modification since the two sink nodes $x_a$, $y_a$ take the top two spots, and since their values are independent of the sets $P,Q$ (by the construction) we cannot use them to distinguish between $P,Q$. We shall add edges $x_a \to u$ in Alice's subgraph and $y_a \to v$ in Bob's subgraph, and self-loops at nodes $u,u',v,v'$. The idea is that this will force one of $u,v$ to be of maximal rank without affecting the fundamental properties of the construction (since the flux transferred from $x_a$ to $u$ is constant). Therefore $v$ can check whether it is the maximum (in which case $P<Q$) and if not, then $u$ must be the maximum (in which case $P>Q$). The full construction is shown in Figure \ref{fig:ranks}. Note the similarity to the construction of Lemma \ref{lemma:denselowerbound}.

We will only work through the details of the construction for the case where each node $x_i$ on the cycle links directly to $u$ or $u'$ (and similarly for the $y_i$ in Bob's half of the network) -- this will give the $O(n)$-bit encoding for sparse graphs, and it can be improved as before to $O(n \log n)$ for dense graphs by the technique of having each $x_i$ link to $O(n)$ intermediate nodes (in which case the algebra becomes quite messy).

For ease of notation, we shall use $p_j=\pi({x_j})$, and $q_j=\pi({y_j})$. Letting $s=\pi(u)+\pi({u'})$ and $t=\pi(v) + \pi({v'})$, we have:
\begin{eqnarray*}
\pi({x_a}) = \frac{1}{2} (\pi(u) + \pi({u'})) &=& \frac{1}{2} s \\
&=& \frac{1}{4} (\pi({x_a}) + (2^m - 1) p_m + \frac{t}{2} + \frac{s}{2}) \\
\Rightarrow s &=& \frac{2}{3} ( \pi({x_a}) + (2^m - 1) p_m + \frac{t}{2}) \\
\Rightarrow \pi({x_a}) &=& \frac{1}{3} (\pi({x_a}) + (2^m - 1) p_m + \frac{t}{2}) \\
&=& \frac{1}{2} ((2^m - 1) p_m + \frac{t}{2}) \\
\end{eqnarray*}
and similarly $\pi({y_a}) = \frac{1}{2} ((2^m - 1) q_m + \frac{s}{2})$. Combining these equations, we get
\begin{eqnarray*}
s+t &=& (2^m - 1) (p_m + q_m) + \frac{1}{2}(s+t) \\
&=& 2 (2^m - 1) (p_m + q_m).
\end{eqnarray*}
Since the sum of all stationary probabilities is 1, this gives
\begin{eqnarray*}
s + t + \pi({x_a}) + \pi({y_a}) + (2^m - 1) (p_m + q_m) &=& 1 \\
\frac{3}{2} (s+t) + (2^m - 1)(p_m + q_m) &=& 1 \\
4 (2^m - 1)(p_m + q_m) &=& 1 \\
p_m + q_m &=& c
\end{eqnarray*}
where $c$ is a constant that depends only on $m=2n$. We can also show that both $p_m,q_m$ are constant, since
\begin{eqnarray*}
s = \pi(u) + \pi({u'}) &=& (2^m - 1) p_m + \frac{t}{2} \\
&=& (2^m - 1) p_m + \frac{1}{2} ((2^m - 1) q_m + \frac{s}{2})
\end{eqnarray*}
which solves to give
\begin{equation}
\label{eqn:s1}
s = \frac{2}{3} (2^m - 1) (2 p_m + q_m).
\end{equation}
Now we also have $p_1 = p_m/2 + \pi({x_a}) / 2 = 2^{m-1} p_m$, so $\pi({x_a}) = (2^m - 1) p_m$ and hence
\begin{equation}
\label{eqn:s2}
s=2 \pi({x_a}) = 2 (2^m - 1) p_m.
\end{equation}
Combining (\ref{eqn:s1}) and (\ref{eqn:s2}) gives $2 (2^m - 1) p_m = \frac{2}{3}(2^m - 1) (2 p_m + q_m)$ and hence $p_m = q_m = c/2$. Therefore both are constants even though they are in different partitions; this means that the flux flowing around both cycles is independent of the sets $P,Q$. Now we can use this to find the value $\pi(u)$:
\begin{eqnarray*}
\pi(u) &=& \frac{1}{2} \sum_{j \in P} p_j + \frac{1}{4} \pi(v) + \frac{1}{2} \pi({x_a}) + \frac{1}{4} \pi(u) \\
\Rightarrow \pi(u) &=& \frac{2}{3} \sum_{j \in P} p_j + \frac{1}{3} \pi(v) + \frac{2}{3} \pi({x_a}) \\
&=& \frac{2}{3} \left( p_m \sum_{j \in P} 2^{m-2j} + \frac{1}{2} \pi(v) + \pi({x_a}) \right) \\
&=& \frac{2}{3} \left( p_m \sum_{j \in P} 2^{m-2j} + \pi({x_a}) +{} \right.\\
&& {}\left. + \frac{1}{3} \left[ q_m \sum_{k \in Q} 2^{m-2k} + \frac{1}{2} \pi(u) + \pi({y_a}) \right] \right).
\end{eqnarray*}
A little manipulation, and recalling that $p_m = q_m = c/2$, gives
\begin{eqnarray*}
\pi(u) &=& \frac{c}{8} \left( 3 \sum_{j \in P} 2^{m-2j} + \sum_{k \in Q} 2^{m-2k} \right) + \frac{c}{2} (2^m - 1) \\
&=& c' \left( 3 \sum_{j \in P} 2^{-2j} + \sum_{k \in Q} 2^{-2k} \right) + \frac{c}{2} (2^m - 1)
\end{eqnarray*}
for some constant $c'$.

Now we just need to check that one of $u,v$ is of maximal rank. The edge from $x_a$ to $u$ ensures that $u$ has higher rank than $u'$, and the self-loops at $u,u',v,v'$ ensure that $u$ has rank at least as high as $x_a$, since $\pi({x_a}) = \frac{c}{2} (2^m - 1) \leq \pi(u)$ (and similarly for the other half of the network). By breaking ties in favour of $u,v$, one of them has maximal rank.

Define $\hat P = \{j \in P ~|~ (|P|-j+1) \in P\}$. Since one of $u,v$ is of maximal rank, Lemma \ref{lemma:max} implies that node $u$ is of maximal rank iff $\hat P > \hat Q$ (wrt to their binary expansions).

Now if some node knows that $u$ is of maximal rank then $\hat P > \hat Q$ otherwise $v$ has maximal rank and so $\hat Q < \hat P$. As shown in the Figure, we can increase the separation factor on the cycle (between successive $x_i$'s) to $O(n)$, and so if Alice knows whether $v$ is of maximal rank, she can solve greater-than on sets of size $O(n \log n)$ bits.
\end{proof}

By the deterministic communication complexity of greater-than we have the following corollary.
\begin{corollary}
Consider any deterministic algorithm that terminates with at least one node knowing if it is of maximal rank. Then at least $\Omega(n \log n)$ bits must be sent over $\Omega(n)$ edges in the network, in the worst case.
\end{corollary}

For randomized algorithms, the situation is somewhat different. The randomized complexity of greater-than is $\Omega(\log n)$.
\begin{corollary}
Consider any algorithm that terminates with at least one node knowing if it is of maximal rank, with probability at least 2/3. Then at least $\Omega(\log n + \log \log n)=\Omega(\log n)$ bits must be sent over $\Omega(n)$ edges in the network, in the worst case.
\end{corollary}

An interesting open problem is to find a distributed deterministic algorithm for computing the maximal node of a Markov chain.

\section{Discussion}
\label{section:conclusion}

We have presented several lower bounds for an interesting problem in distributed computing, where the structure of the communication network is the input to the function to be computed. Our technique is to embed an instance of some two-party version of the problem into a network, and by appealing to the linear array conjecture, lifting the two-party result onto a result concerning the total communication of a distributed algorithm. We discussed that strengthening our results is likely to require a different lifting technique, as the linear array lifting only lets us account for the flow of data across a linear number of edges in $n$, even though there may be $O(n^2)$ edges present. Finding a better lifting technique appears to be a general problem in proving good lower bounds for distributed computing problems.

In considering worst-case complexity we have neglected the graph-theoretic properties of $G$ -- it would be useful to know how the communication complexity of the problems is altered by restricting $G$ to say, graphs of high conductance (in particular, this would seem to reduce the effectiveness of the lifting technique because a graph of high conductance could not contain two dense graphs separated by a long string of edges, as this would then resemble a `barbell graph').

Some of our lower bound reductions involving greater-than suggest that randomization may help. So far, we have been unable to confirm this but it would certainly seem natural, given the random walk interpretation of the problem.

%% file: cycle1.pstex_t
\begin{picture}(0,0)%
\includegraphics{cycle1.pstex}%
\end{picture}%
\setlength{\unitlength}{1342sp}%
\begingroup\makeatletter\ifx\SetFigFont\undefined%
\gdef\SetFigFont#1#2#3#4#5{%
  \reset@font\fontsize{#1}{#2pt}%
  \fontfamily{#3}\fontseries{#4}\fontshape{#5}%
  \selectfont}%
\fi\endgroup%
\begin{picture}(10040,9849)(2781,-8773)
\put(9601,-8236){\makebox(0,0)[b]{\smash{\SetFigFont{6}{7.2}{\rmdefault}{\mddefault}{\updefault}{\color[rgb]{0,0,0}$u'$}%
}}}
\put(9601,164){\makebox(0,0)[b]{\smash{\SetFigFont{6}{7.2}{\rmdefault}{\mddefault}{\updefault}{\color[rgb]{0,0,0}$u$}%
}}}
\put(7801,-4036){\makebox(0,0)[b]{\smash{\SetFigFont{6}{7.2}{\rmdefault}{\mddefault}{\updefault}{\color[rgb]{0,0,0}$x_3$}%
}}}
\put(5401,-4036){\makebox(0,0)[b]{\smash{\SetFigFont{6}{7.2}{\rmdefault}{\mddefault}{\updefault}{\color[rgb]{0,0,0}$x_a$}%
}}}
\put(3001,-4036){\makebox(0,0)[b]{\smash{\SetFigFont{6}{7.2}{\rmdefault}{\mddefault}{\updefault}{\color[rgb]{0,0,0}$x_1$}%
}}}
\put(12601,-8236){\makebox(0,0)[b]{\smash{\SetFigFont{6}{7.2}{\rmdefault}{\mddefault}{\updefault}{\color[rgb]{0,0,0}$v'$}%
}}}
\put(12601,164){\makebox(0,0)[b]{\smash{\SetFigFont{6}{7.2}{\rmdefault}{\mddefault}{\updefault}{\color[rgb]{0,0,0}$v$}%
}}}
\put(5401,-1636){\makebox(0,0)[b]{\smash{\SetFigFont{6}{7.2}{\rmdefault}{\mddefault}{\updefault}{\color[rgb]{0,0,0}$x_2$}%
}}}
\put(5401,-6436){\makebox(0,0)[b]{\smash{\SetFigFont{6}{7.2}{\rmdefault}{\mddefault}{\updefault}{\color[rgb]{0,0,0}$x_4$}%
}}}
\put(12526,839){\makebox(0,0)[b]{\smash{\SetFigFont{6}{7.2}{\rmdefault}{\mddefault}{\updefault}{\color[rgb]{0,0,0}$\mathbf{Bob}$}%
}}}
\put(8251,839){\makebox(0,0)[b]{\smash{\SetFigFont{6}{7.2}{\rmdefault}{\mddefault}{\updefault}{\color[rgb]{0,0,0}$\mathbf{Alice}$}%
}}}
\end{picture}

%% file: cycledense.pstex_t
\begin{picture}(0,0)%
\includegraphics{cycledense.pstex}%
\end{picture}%
\setlength{\unitlength}{1579sp}%
\begingroup\makeatletter\ifx\SetFigFont\undefined%
\gdef\SetFigFont#1#2#3#4#5{%
  \reset@font\fontsize{#1}{#2pt}%
  \fontfamily{#3}\fontseries{#4}\fontshape{#5}%
  \selectfont}%
\fi\endgroup%
\begin{picture}(11946,9789)(1201,-9133)
\put(3451,-5836){\makebox(0,0)[b]{\smash{\SetFigFont{7}{8.4}{\rmdefault}{\mddefault}{\updefault}{\color[rgb]{0,0,0}$x_1$}%
}}}
\put(3451,-2536){\makebox(0,0)[b]{\smash{\SetFigFont{7}{8.4}{\rmdefault}{\mddefault}{\updefault}{\color[rgb]{0,0,0}$x_2$}%
}}}
\put(6376,-2011){\makebox(0,0)[b]{\smash{\SetFigFont{7}{8.4}{\rmdefault}{\mddefault}{\updefault}{\color[rgb]{0,0,0}$x_3$}%
}}}
\put(7801,-4036){\makebox(0,0)[b]{\smash{\SetFigFont{7}{8.4}{\rmdefault}{\mddefault}{\updefault}{\color[rgb]{0,0,0}$x_4$}%
}}}
\put(6451,-6361){\makebox(0,0)[b]{\smash{\SetFigFont{7}{8.4}{\rmdefault}{\mddefault}{\updefault}{\color[rgb]{0,0,0}$x_5$}%
}}}
\put(7801,-8236){\makebox(0,0)[b]{\smash{\SetFigFont{7}{8.4}{\rmdefault}{\mddefault}{\updefault}{\color[rgb]{0,0,0}$y'_1$}%
}}}
\put(9001,-8236){\makebox(0,0)[b]{\smash{\SetFigFont{7}{8.4}{\rmdefault}{\mddefault}{\updefault}{\color[rgb]{0,0,0}$y'_2$}%
}}}
\put(10201,-8236){\makebox(0,0)[b]{\smash{\SetFigFont{7}{8.4}{\rmdefault}{\mddefault}{\updefault}{\color[rgb]{0,0,0}$y'_3$}%
}}}
\put(11401,-8236){\makebox(0,0)[b]{\smash{\SetFigFont{7}{8.4}{\rmdefault}{\mddefault}{\updefault}{\color[rgb]{0,0,0}$y'_4$}%
}}}
\put(12601,-8236){\makebox(0,0)[b]{\smash{\SetFigFont{7}{8.4}{\rmdefault}{\mddefault}{\updefault}{\color[rgb]{0,0,0}$y'_5$}%
}}}
\put(9001,-436){\makebox(0,0)[b]{\smash{\SetFigFont{7}{8.4}{\rmdefault}{\mddefault}{\updefault}{\color[rgb]{0,0,0}$y_2$}%
}}}
\put(7726,-436){\makebox(0,0)[b]{\smash{\SetFigFont{7}{8.4}{\rmdefault}{\mddefault}{\updefault}{\color[rgb]{0,0,0}$y_1$}%
}}}
\put(10201,-361){\makebox(0,0)[b]{\smash{\SetFigFont{7}{8.4}{\rmdefault}{\mddefault}{\updefault}{\color[rgb]{0,0,0}$y_3$}%
}}}
\put(11326,-361){\makebox(0,0)[b]{\smash{\SetFigFont{7}{8.4}{\rmdefault}{\mddefault}{\updefault}{\color[rgb]{0,0,0}$y_4$}%
}}}
\put(12601,-436){\makebox(0,0)[b]{\smash{\SetFigFont{7}{8.4}{\rmdefault}{\mddefault}{\updefault}{\color[rgb]{0,0,0}$y_5$}%
}}}
\put(1201,-2911){\makebox(0,0)[b]{\smash{\SetFigFont{7}{8.4}{\rmdefault}{\mddefault}{\updefault}{\color[rgb]{0,0,0}To $y'_1$ .. $y'_5$}%
}}}
\put(1276,-5311){\makebox(0,0)[b]{\smash{\SetFigFont{7}{8.4}{\rmdefault}{\mddefault}{\updefault}{\color[rgb]{0,0,0}To $y_1$ .. $y_5$}%
}}}
\put(5401,-4036){\makebox(0,0)[b]{\smash{\SetFigFont{7}{8.4}{\rmdefault}{\mddefault}{\updefault}{\color[rgb]{0,0,0}$s$}%
}}}
\put(9076,464){\makebox(0,0)[b]{\smash{\SetFigFont{7}{8.4}{\rmdefault}{\mddefault}{\updefault}{\color[rgb]{0,0,0}To $s$}%
}}}
\put(9151,-9061){\makebox(0,0)[b]{\smash{\SetFigFont{7}{8.4}{\rmdefault}{\mddefault}{\updefault}{\color[rgb]{0,0,0}To $s$}%
}}}
\put(5176,-1111){\makebox(0,0)[b]{\smash{\SetFigFont{7}{8.4}{\rmdefault}{\mddefault}{\updefault}{\color[rgb]{0,0,0}$\mathbf{Alice}$}%
}}}
\put(6451,-61){\makebox(0,0)[b]{\smash{\SetFigFont{7}{8.4}{\rmdefault}{\mddefault}{\updefault}{\color[rgb]{0,0,0}$\mathbf{Bob}$}%
}}}
\end{picture}

%% file: rank.pstex_t
\begin{picture}(0,0)%
\includegraphics{rank.pstex}%
\end{picture}%
\setlength{\unitlength}{1776sp}%
\begingroup\makeatletter\ifx\SetFigFont\undefined%
\gdef\SetFigFont#1#2#3#4#5{%
  \reset@font\fontsize{#1}{#2pt}%
  \fontfamily{#3}\fontseries{#4}\fontshape{#5}%
  \selectfont}%
\fi\endgroup%
\begin{picture}(12580,8049)(-64,-7423)
\put(10576,-5536){\makebox(0,0)[b]{\smash{\SetFigFont{8}{9.6}{\rmdefault}{\mddefault}{\updefault}{\color[rgb]{0,0,0}$y_6$}%
}}}
\put(9976,-6136){\makebox(0,0)[b]{\smash{\SetFigFont{8}{9.6}{\rmdefault}{\mddefault}{\updefault}{\color[rgb]{0,0,0}$b'_3$}%
}}}
\put(9976,-4936){\makebox(0,0)[b]{\smash{\SetFigFont{8}{9.6}{\rmdefault}{\mddefault}{\updefault}{\color[rgb]{0,0,0}$b_3$}%
}}}
\put(9376,-5536){\makebox(0,0)[b]{\smash{\SetFigFont{8}{9.6}{\rmdefault}{\mddefault}{\updefault}{\color[rgb]{0,0,0}$y_5$}%
}}}
\put(8026,-2986){\rotatebox{90.0}{\makebox(0,0)[b]{\smash{\SetFigFont{8}{9.6}{\rmdefault}{\mddefault}{\updefault}{\color[rgb]{0,0,0}$y_3$}%
}}}}
\put(8026,-4186){\rotatebox{90.0}{\makebox(0,0)[b]{\smash{\SetFigFont{8}{9.6}{\rmdefault}{\mddefault}{\updefault}{\color[rgb]{0,0,0}$y_4$}%
}}}}
\put(7426,-3586){\rotatebox{90.0}{\makebox(0,0)[b]{\smash{\SetFigFont{8}{9.6}{\rmdefault}{\mddefault}{\updefault}{\color[rgb]{0,0,0}$b'_2$}%
}}}}
\put(8626,-3586){\rotatebox{90.0}{\makebox(0,0)[b]{\smash{\SetFigFont{8}{9.6}{\rmdefault}{\mddefault}{\updefault}{\color[rgb]{0,0,0}$b_2$}%
}}}}
\put(10426,-1636){\makebox(0,0)[b]{\smash{\SetFigFont{8}{9.6}{\rmdefault}{\mddefault}{\updefault}{\color[rgb]{0,0,0}$y_1$}%
}}}
\put(9226,-1636){\makebox(0,0)[b]{\smash{\SetFigFont{8}{9.6}{\rmdefault}{\mddefault}{\updefault}{\color[rgb]{0,0,0}$y_2$}%
}}}
\put(9826,-1036){\makebox(0,0)[b]{\smash{\SetFigFont{8}{9.6}{\rmdefault}{\mddefault}{\updefault}{\color[rgb]{0,0,0}$b'_1$}%
}}}
\put(9826,-2236){\makebox(0,0)[b]{\smash{\SetFigFont{8}{9.6}{\rmdefault}{\mddefault}{\updefault}{\color[rgb]{0,0,0}$b_1$}%
}}}
\put(11476,-2986){\rotatebox{270.0}{\makebox(0,0)[b]{\smash{\SetFigFont{8}{9.6}{\rmdefault}{\mddefault}{\updefault}{\color[rgb]{0,0,0}$y_0$}%
}}}}
\put(11476,-4186){\rotatebox{270.0}{\makebox(0,0)[b]{\smash{\SetFigFont{8}{9.6}{\rmdefault}{\mddefault}{\updefault}{\color[rgb]{0,0,0}$y_7$}%
}}}}
\put(9751,-3211){\makebox(0,0)[b]{\smash{\SetFigFont{8}{9.6}{\rmdefault}{\mddefault}{\updefault}{\color[rgb]{0,0,0}$t$}%
}}}
\put(12301,-2461){\rotatebox{270.0}{\makebox(0,0)[b]{\smash{\SetFigFont{8}{9.6}{\rmdefault}{\mddefault}{\updefault}{\color[rgb]{0,0,0}To $b'_1 .. b'_n$}%
}}}}
\put(1951,-5536){\makebox(0,0)[b]{\smash{\SetFigFont{8}{9.6}{\rmdefault}{\mddefault}{\updefault}{\color[rgb]{0,0,0}$x_6$}%
}}}
\put(2551,-6136){\makebox(0,0)[b]{\smash{\SetFigFont{8}{9.6}{\rmdefault}{\mddefault}{\updefault}{\color[rgb]{0,0,0}$a'_3$}%
}}}
\put(2551,-4936){\makebox(0,0)[b]{\smash{\SetFigFont{8}{9.6}{\rmdefault}{\mddefault}{\updefault}{\color[rgb]{0,0,0}$a_3$}%
}}}
\put(3151,-5536){\makebox(0,0)[b]{\smash{\SetFigFont{8}{9.6}{\rmdefault}{\mddefault}{\updefault}{\color[rgb]{0,0,0}$x_5$}%
}}}
\put(4501,-2986){\rotatebox{270.0}{\makebox(0,0)[b]{\smash{\SetFigFont{8}{9.6}{\rmdefault}{\mddefault}{\updefault}{\color[rgb]{0,0,0}$x_3$}%
}}}}
\put(4501,-4186){\rotatebox{270.0}{\makebox(0,0)[b]{\smash{\SetFigFont{8}{9.6}{\rmdefault}{\mddefault}{\updefault}{\color[rgb]{0,0,0}$x_4$}%
}}}}
\put(5101,-3586){\rotatebox{270.0}{\makebox(0,0)[b]{\smash{\SetFigFont{8}{9.6}{\rmdefault}{\mddefault}{\updefault}{\color[rgb]{0,0,0}$a'_2$}%
}}}}
\put(3901,-3586){\rotatebox{270.0}{\makebox(0,0)[b]{\smash{\SetFigFont{8}{9.6}{\rmdefault}{\mddefault}{\updefault}{\color[rgb]{0,0,0}$a_2$}%
}}}}
\put(2101,-1636){\makebox(0,0)[b]{\smash{\SetFigFont{8}{9.6}{\rmdefault}{\mddefault}{\updefault}{\color[rgb]{0,0,0}$x_1$}%
}}}
\put(3301,-1636){\makebox(0,0)[b]{\smash{\SetFigFont{8}{9.6}{\rmdefault}{\mddefault}{\updefault}{\color[rgb]{0,0,0}$x_2$}%
}}}
\put(1051,-2986){\rotatebox{90.0}{\makebox(0,0)[b]{\smash{\SetFigFont{8}{9.6}{\rmdefault}{\mddefault}{\updefault}{\color[rgb]{0,0,0}$x_0$}%
}}}}
\put(1051,-4186){\rotatebox{90.0}{\makebox(0,0)[b]{\smash{\SetFigFont{8}{9.6}{\rmdefault}{\mddefault}{\updefault}{\color[rgb]{0,0,0}$x_7$}%
}}}}
\put(2776,-3211){\makebox(0,0)[b]{\smash{\SetFigFont{8}{9.6}{\rmdefault}{\mddefault}{\updefault}{\color[rgb]{0,0,0}$s$}%
}}}
\put(2701,-2236){\makebox(0,0)[b]{\smash{\SetFigFont{8}{9.6}{\rmdefault}{\mddefault}{\updefault}{\color[rgb]{0,0,0}$a_1$}%
}}}
\put(2701,-1036){\makebox(0,0)[b]{\smash{\SetFigFont{8}{9.6}{\rmdefault}{\mddefault}{\updefault}{\color[rgb]{0,0,0}$a'_1$}%
}}}
\put(151,-2686){\rotatebox{90.0}{\makebox(0,0)[b]{\smash{\SetFigFont{8}{9.6}{\rmdefault}{\mddefault}{\updefault}{\color[rgb]{0,0,0}To $a'_1 .. a'_n$}%
}}}}
\put(10801,-4036){\rotatebox{90.0}{\makebox(0,0)[b]{\smash{\SetFigFont{8}{9.6}{\rmdefault}{\mddefault}{\updefault}{\color[rgb]{0,0,0}To $b_1 .. b_n$}%
}}}}
\put(6901,-2761){\makebox(0,0)[b]{\smash{\SetFigFont{8}{9.6}{\rmdefault}{\mddefault}{\updefault}{\color[rgb]{0,0,0}To $t$}%
}}}
\put(10651,-736){\makebox(0,0)[b]{\smash{\SetFigFont{8}{9.6}{\rmdefault}{\mddefault}{\updefault}{\color[rgb]{0,0,0}To $t$}%
}}}
\put(9076,-6436){\makebox(0,0)[b]{\smash{\SetFigFont{8}{9.6}{\rmdefault}{\mddefault}{\updefault}{\color[rgb]{0,0,0}To $t$}%
}}}
\put(1801,-3961){\rotatebox{270.0}{\makebox(0,0)[b]{\smash{\SetFigFont{8}{9.6}{\rmdefault}{\mddefault}{\updefault}{\color[rgb]{0,0,0}To $a_1 .. a_n$}%
}}}}
\put(1501,-6286){\makebox(0,0)[b]{\smash{\SetFigFont{8}{9.6}{\rmdefault}{\mddefault}{\updefault}{\color[rgb]{0,0,0}To $s$}%
}}}
\put(3601,-811){\makebox(0,0)[b]{\smash{\SetFigFont{8}{9.6}{\rmdefault}{\mddefault}{\updefault}{\color[rgb]{0,0,0}To $s$}%
}}}
\put(5476,-4636){\makebox(0,0)[b]{\smash{\SetFigFont{8}{9.6}{\rmdefault}{\mddefault}{\updefault}{\color[rgb]{0,0,0}To $s$}%
}}}
\put(4951, 89){\makebox(0,0)[b]{\smash{\SetFigFont{8}{9.6}{\rmdefault}{\mddefault}{\updefault}{\color[rgb]{0,0,0}$\mathbf{Alice}$}%
}}}
\put(7426, 89){\makebox(0,0)[b]{\smash{\SetFigFont{8}{9.6}{\rmdefault}{\mddefault}{\updefault}{\color[rgb]{0,0,0}$\mathbf{Bob}$}%
}}}
\end{picture}

%% file: ranks.pstex_t
\begin{picture}(0,0)%
\includegraphics{ranks.pstex}%
\end{picture}%
\setlength{\unitlength}{1302sp}%
\begingroup\makeatletter\ifx\SetFigFont\undefined%
\gdef\SetFigFont#1#2#3#4#5{%
  \reset@font\fontsize{#1}{#2pt}%
  \fontfamily{#3}\fontseries{#4}\fontshape{#5}%
  \selectfont}%
\fi\endgroup%
\begin{picture}(24440,9672)(2781,-7867)
\put(16576,-2686){\makebox(0,0)[b]{\smash{\SetFigFont{6}{7.2}{\rmdefault}{\mddefault}{\updefault}{\color[rgb]{0,0,0}$v$}%
}}}
\put(16576,-5236){\makebox(0,0)[b]{\smash{\SetFigFont{6}{7.2}{\rmdefault}{\mddefault}{\updefault}{\color[rgb]{0,0,0}$v'$}%
}}}
\put(22201,-4036){\makebox(0,0)[b]{\smash{\SetFigFont{6}{7.2}{\rmdefault}{\mddefault}{\updefault}{\color[rgb]{0,0,0}$y_3$}%
}}}
\put(27001,-4036){\makebox(0,0)[b]{\smash{\SetFigFont{6}{7.2}{\rmdefault}{\mddefault}{\updefault}{\color[rgb]{0,0,0}$y_1$}%
}}}
\put(24601,-1636){\makebox(0,0)[b]{\smash{\SetFigFont{6}{7.2}{\rmdefault}{\mddefault}{\updefault}{\color[rgb]{0,0,0}$y_2$}%
}}}
\put(24601,-6436){\makebox(0,0)[b]{\smash{\SetFigFont{6}{7.2}{\rmdefault}{\mddefault}{\updefault}{\color[rgb]{0,0,0}$y_4$}%
}}}
\put(21976,164){\makebox(0,0)[b]{\smash{\SetFigFont{6}{7.2}{\rmdefault}{\mddefault}{\updefault}{\color[rgb]{0,0,0}$y_a$}%
}}}
\put(26401,-5161){\makebox(0,0)[b]{\smash{\SetFigFont{6}{7.2}{\rmdefault}{\mddefault}{\updefault}{\color[rgb]{0,0,0}.}%
}}}
\put(26251,-5311){\makebox(0,0)[b]{\smash{\SetFigFont{6}{7.2}{\rmdefault}{\mddefault}{\updefault}{\color[rgb]{0,0,0}.}%
}}}
\put(26101,-5461){\makebox(0,0)[b]{\smash{\SetFigFont{6}{7.2}{\rmdefault}{\mddefault}{\updefault}{\color[rgb]{0,0,0}.}%
}}}
\put(19426,1514){\makebox(0,0)[b]{\smash{\SetFigFont{6}{7.2}{\rmdefault}{\mddefault}{\updefault}{\color[rgb]{0,0,0}$\mathbf{Bob}$}%
}}}
\put(9976,1589){\makebox(0,0)[b]{\smash{\SetFigFont{6}{7.2}{\rmdefault}{\mddefault}{\updefault}{\color[rgb]{0,0,0}$\mathbf{Alice}$}%
}}}
\put(7801,-4036){\makebox(0,0)[b]{\smash{\SetFigFont{6}{7.2}{\rmdefault}{\mddefault}{\updefault}{\color[rgb]{0,0,0}$x_3$}%
}}}
\put(3001,-4036){\makebox(0,0)[b]{\smash{\SetFigFont{6}{7.2}{\rmdefault}{\mddefault}{\updefault}{\color[rgb]{0,0,0}$x_1$}%
}}}
\put(5401,-1636){\makebox(0,0)[b]{\smash{\SetFigFont{6}{7.2}{\rmdefault}{\mddefault}{\updefault}{\color[rgb]{0,0,0}$x_2$}%
}}}
\put(5401,-6436){\makebox(0,0)[b]{\smash{\SetFigFont{6}{7.2}{\rmdefault}{\mddefault}{\updefault}{\color[rgb]{0,0,0}$x_4$}%
}}}
\put(13426,-2686){\makebox(0,0)[b]{\smash{\SetFigFont{6}{7.2}{\rmdefault}{\mddefault}{\updefault}{\color[rgb]{0,0,0}$u$}%
}}}
\put(13426,-5236){\makebox(0,0)[b]{\smash{\SetFigFont{6}{7.2}{\rmdefault}{\mddefault}{\updefault}{\color[rgb]{0,0,0}$u'$}%
}}}
\put(8026,164){\makebox(0,0)[b]{\smash{\SetFigFont{6}{7.2}{\rmdefault}{\mddefault}{\updefault}{\color[rgb]{0,0,0}$x_a$}%
}}}
\put(3601,-5161){\makebox(0,0)[b]{\smash{\SetFigFont{6}{7.2}{\rmdefault}{\mddefault}{\updefault}{\color[rgb]{0,0,0}.}%
}}}
\put(3751,-5311){\makebox(0,0)[b]{\smash{\SetFigFont{6}{7.2}{\rmdefault}{\mddefault}{\updefault}{\color[rgb]{0,0,0}.}%
}}}
\put(3901,-5461){\makebox(0,0)[b]{\smash{\SetFigFont{6}{7.2}{\rmdefault}{\mddefault}{\updefault}{\color[rgb]{0,0,0}.}%
}}}
\put(11626,-1711){\makebox(0,0)[b]{\smash{\SetFigFont{6}{7.2}{\rmdefault}{\mddefault}{\updefault}{\color[rgb]{0,0,0}$z_1$}%
}}}
\put(11626,-2311){\makebox(0,0)[b]{\smash{\SetFigFont{6}{7.2}{\rmdefault}{\mddefault}{\updefault}{\color[rgb]{0,0,0}$z_2$}%
}}}
\put(11626,-4561){\makebox(0,0)[b]{\smash{\SetFigFont{6}{7.2}{\rmdefault}{\mddefault}{\updefault}{\color[rgb]{0,0,0}$z'_1$}%
}}}
\put(11626,-5161){\makebox(0,0)[b]{\smash{\SetFigFont{6}{7.2}{\rmdefault}{\mddefault}{\updefault}{\color[rgb]{0,0,0}$z'_2$}%
}}}
\put(11626,-3136){\makebox(0,0)[b]{\smash{\SetFigFont{6}{7.2}{\rmdefault}{\mddefault}{\updefault}{\color[rgb]{0,0,0}$\vdots$}%
}}}
\put(11626,-5986){\makebox(0,0)[b]{\smash{\SetFigFont{6}{7.2}{\rmdefault}{\mddefault}{\updefault}{\color[rgb]{0,0,0}$\vdots$}%
}}}
\put(18376,-5986){\makebox(0,0)[b]{\smash{\SetFigFont{6}{7.2}{\rmdefault}{\mddefault}{\updefault}{\color[rgb]{0,0,0}$\vdots$}%
}}}
\put(18376,-3136){\makebox(0,0)[b]{\smash{\SetFigFont{6}{7.2}{\rmdefault}{\mddefault}{\updefault}{\color[rgb]{0,0,0}$\vdots$}%
}}}
\put(18376,-1711){\makebox(0,0)[b]{\smash{\SetFigFont{6}{7.2}{\rmdefault}{\mddefault}{\updefault}{\color[rgb]{0,0,0}$z_1$}%
}}}
\put(18376,-2311){\makebox(0,0)[b]{\smash{\SetFigFont{6}{7.2}{\rmdefault}{\mddefault}{\updefault}{\color[rgb]{0,0,0}$z_2$}%
}}}
\put(18376,-4561){\makebox(0,0)[b]{\smash{\SetFigFont{6}{7.2}{\rmdefault}{\mddefault}{\updefault}{\color[rgb]{0,0,0}$z'_1$}%
}}}
\put(18376,-5161){\makebox(0,0)[b]{\smash{\SetFigFont{6}{7.2}{\rmdefault}{\mddefault}{\updefault}{\color[rgb]{0,0,0}$z'_2$}%
}}}
\put(5476,1289){\makebox(0,0)[b]{\smash{\SetFigFont{6}{7.2}{\rmdefault}{\mddefault}{\updefault}{\color[rgb]{0,0,0}These edges encode a $O(log n)$-element set}%
}}}
\put(6826,-7786){\makebox(0,0)[b]{\smash{\SetFigFont{6}{7.2}{\rmdefault}{\mddefault}{\updefault}{\color[rgb]{0,0,0}Odd nodes on the cycle do not contribute anything to $\pi_u$}%
}}}
\put(11626,-3736){\makebox(0,0)[b]{\smash{\SetFigFont{6}{7.2}{\rmdefault}{\mddefault}{\updefault}{\color[rgb]{0,0,0}$z_n$}%
}}}
\put(11626,-6586){\makebox(0,0)[b]{\smash{\SetFigFont{6}{7.2}{\rmdefault}{\mddefault}{\updefault}{\color[rgb]{0,0,0}$z'_n$}%
}}}
\put(18376,-3736){\makebox(0,0)[b]{\smash{\SetFigFont{6}{7.2}{\rmdefault}{\mddefault}{\updefault}{\color[rgb]{0,0,0}$z_n$}%
}}}
\put(18376,-6586){\makebox(0,0)[b]{\smash{\SetFigFont{6}{7.2}{\rmdefault}{\mddefault}{\updefault}{\color[rgb]{0,0,0}$z'_n$}%
}}}
\end{picture}